\documentclass[%
 reprint,
 superscriptaddress,
 amsmath,amssymb,prx,
]{revtex4-2}

\usepackage{graphicx}% Include figure files
\usepackage{dcolumn}% Align table columns on decimal point
\usepackage{bm}% bold math
\usepackage{hyperref}% add hypertext capabilities
\usepackage{xcolor}
\begin{document}

\title{The Laser-Diode Heated Floating Zone Method for Automated Optimal Synthesis of Refractory Oxides and Alloys}

\author{Alemayehu S. Admasu}
\email{admasu@physics.rutgers.edu/aadmasu@ford.com}
\author{Durga Sankar Vavilapalli}
\affiliation{Rutgers Center for Emergent Materials and Department of Physics and Astronomy, Rutgers University, Piscataway, New Jersey 08854, USA}

\date{\today}

\begin{abstract}
We perform an experimental investigation of the synthesis of complex materials using the Laser-diode heated floating zone method (L-FZ). We will briefly introduce the Infrared-heated floating zone method of bulk crystal growth and then delve into recent advances in using a Laser-diode heated floating zone method. We demonstrate L-FZ for the growth of large high-quality single crystalline samples of  of the n=2 Ruddlesden–Popper homologous series $RE_2SrAl_2O_7$, RE = Nd, Sm, Eu, Gd, Tb, Dy with physically interesting properties (e.g spin-ice) as well as the incongruently melting stuffed-tridymite type oxide $BaCoSiO_4$. We conclude with a summary of the results and future research directions in automating crystal growth, which will open access into the synthesis and characterization of previously unstudied class of materials such as refractory complex oxides and alloys.

\end{abstract}

\maketitle

\section{INTRODUCTION}

It's popularly known in the material science and physics community: “Who dominates materials dominates technology” \cite{elsevier_crystal_growth}. And the ability to dominate materials discovery and make scientific advances thereof requires an in-depth knowledge in the technique and technology of crystal growth, especially of single crystals – which is a core discipline in its own right that saw an impetus in research and development since the invention of the transistor in 1950s and resulting need for high purity semiconductor single crystals. \\

Material discovery proper occupies the central part of material science studies. It includes the investigation of the structure of crystals (among other systems such as thin films, heterostructures, amorphous solid etc.), their synthesis, chemistry and physics in turn contributing to advances in these and other areas of science. Specifically, experimental and theoretical research of crystal physics deals mainly with the electrical, optical, and mechanical properties of crystals closely related to their structure and symmetry and on the analysis of laws defining their general physical properties. \\

Crystals at the microscopic scale are ideally a perfect ordered arrangement of atoms or molecules. The microscopic regularity can be observed in various diffraction patterns. In addition, the regular arrangement of atoms brings about symmetry in crystalline shape/morphology (polyhedral, dendrite etc.) which is also controlled by mechanisms of the growth dynamics. Crystal growth encompasses not only the synthesis of crystals under controlled conditions and the experimental and theoretical investigation of crystallization processes but also crystal characterization using various techniques for both scientific purposes and industrial applications. Crystal growth technology is mainly driven by the need for applications and in search of novel physical phenomena.  \\

In the last few decades, the major industrial applications have been in the fields of electronic, optical and magnetic materials. To name a few, the groups of compounds that have received the most interest from solid-state physicists include: high-temperature and iron-based superconductors  ($La_{2 - x}Sr_xCuO_4$, $YBa_2Cu_3O_7$, $BaFe_2As_2$, $FeSe$), colossal magnetoresistance in manganese oxides  ($La_{1 - x} (Sr,Ca)_xMnO_3$), geometrically frustrated magnetism oxides ($Dy_2Ti_2O_7$), ferroelectric magnets or “multiferroics” ($TbMnO_3$, $BiFeO_3$), topological insulators ($Bi_2Se_3$, $SmB_6$) etc., see \cite{Dong2015} for more examples. It’s easy to notice that most of these compounds share commonalities: most belong to a typical structural group called perovskites, most are ternary compounds/pseudo-ternanry solid solutions, and except for $YBa_2Cu_3O_7$, all others have been known to exist as stable phases before hence in a sense were re-discovered with in this context.  \\

\begin{figure}[h]%
    \centering
    \includegraphics[width=0.3\textwidth]{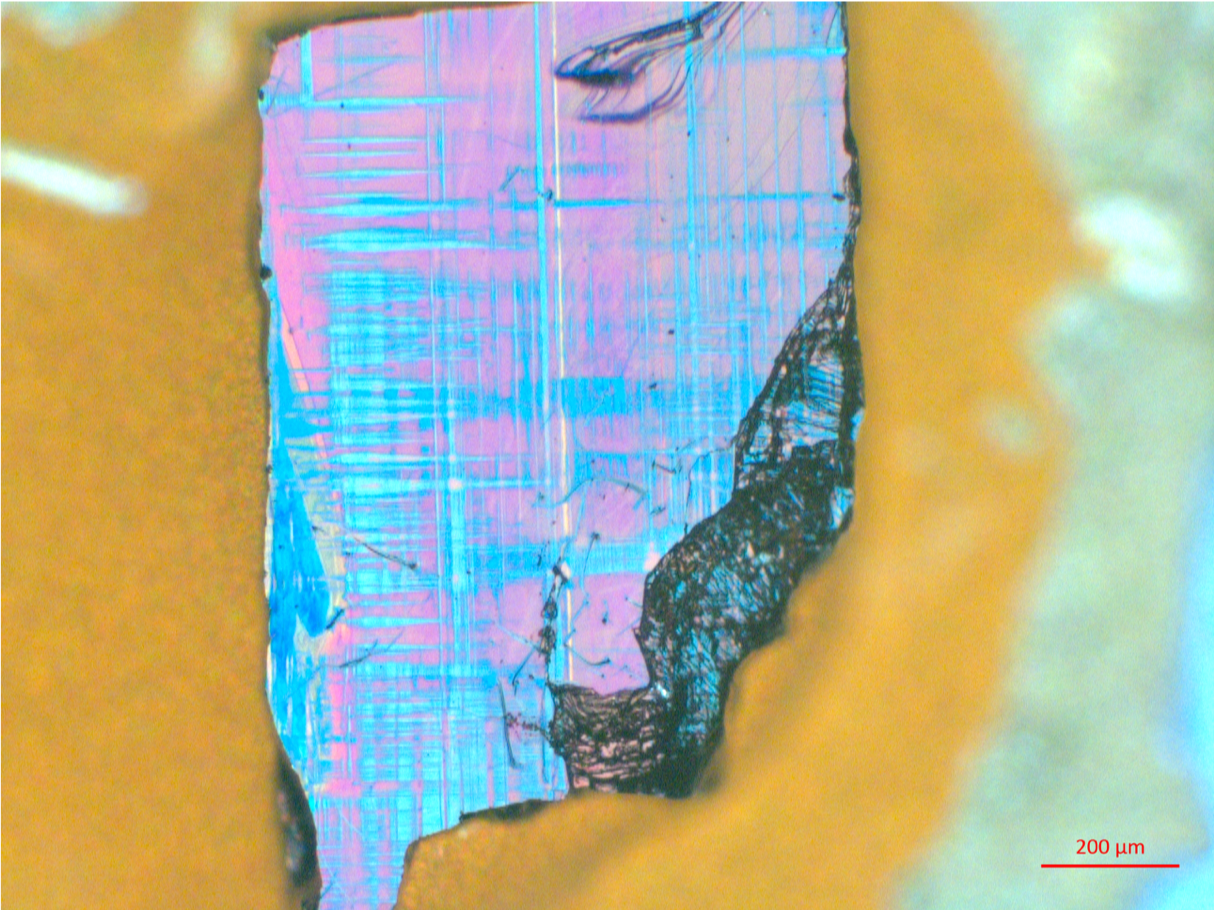} %scale=1.5
    \caption{A cleaved single crystal of $(Pr,Ca,Sr)MnO_3$ under a reflection-Polarized Optical Microscope. \cite{{labfigures}}}%
    \label{fig:PCSMO_cleaved_crystal}%
\end{figure}

For the purposes of this review it is worthy to mention that the crystal growth activity \cite{Feigelson2015}  can be divided into the following types with increasing degree of its difficulty and exploratory nature: (1) synthesis of known compounds with recipes from the literature, (2) synthesis of larger and/or higher quality single crystals (3)  first single crystal growth of already known compounds (4) single crystals synthesis of yet unknown compounds. In this review, we will mainly be concerned with categories 1-3.

%%%%%%%%%%%%%%%%%%%%%%%%%%%%%%%%%%%
\section{Bulk Crystal Growth by Floating Zone Method}

The earliest accounts of crystal growth methods indicate mimicking the ways already observed in nature i.e suspending small `seed' crystals of the to be grown compound in a flux or supersaturated solution resulting in beautiful but small and imperfect crystals, usually containing inclusions, impurities and stacking faults in the crystal lattices \cite{Tatarchenko2003}. To produce larger crystals, `bulk-growth' techniques were invented such as Czochralski (1917), and  Bridgman–Stockbarger (1925) in which the molten state of the required material was cooled slowly through its melting point either by a slow, controlled, pulling of a seed crystal from the melt or by using temperature gradient across the crucible.  \\

The historical development of crystal growth processes is also good reflection of the study and importance of capillarity, surface tension and wetting, stability analysis, shape control, crucible - free concepts etc. as applied in the synthesis controlling mechanisms.  The next prominent crystal bulk crystal growth method to be invented, originally used for material purification (zone refining), was the floating zone (FZ) technique \cite{Theuerer1952} – which is crucible free. Being a purely capillary-based technique, as the melt zone is supported entirely by surface tension, controlling the shape of the crystal or the growth dynamics could be difficult.  A lot of research and development effort have suggested methods to optimize the growth process one of which is the Laser-diode heated FZ that aims to stabilize the growth process using laser radiation at the same time achieving high enough power for growth of high-temperature melting refractory compounds and alloys as we will explain later.    \\

While the technology of growth of commercial bulk crystals is largely based on melt crystallization (the density of material is comparable to the crystal in the melt than the vapor phase or of molecules in a solution, hence making growth rate from the melt relatively faster than other methods), vapor-phase crystallization techniques such as the chemical vapor transport method have also become a wider choice for compounds not primarily suitable for growth by the above methods e.g. transition metal dichalcogenides,  or for which large crystals are not necessary \cite{Jin,Chen}.  \\
In the following sections, we will introduce two main techniques used in the current research of the floating zone method along with the compounds of interest that were studied for novel physical phenomena and characterization.  

% \subsection{}

\subsubsection{Infrared Optical Floating Zone Method}

The (infra-red) floating zone technique uses a molten zone established between two rods by halogen/xenon lamps and ellipsoidal mirrors as a heat source, the two rods being a polycrystalline feedrod at the top and a seed rod/crystal at the bottom. The equilibrium shapes and stability of the floating zone melt involve material characteristics of the melt, growth atmosphere, temperature gradients and other growth complexities that may or may not be monitored during the process. A wide range of high quality and relatively large single crystals (usually on the order of few mm in diameter and few cm long) materials, from metals and semiconductors to oxides can be grown by this technique.  \\

\begin{figure}[h]%
    \centering
    \includegraphics[width=0.3\textwidth]{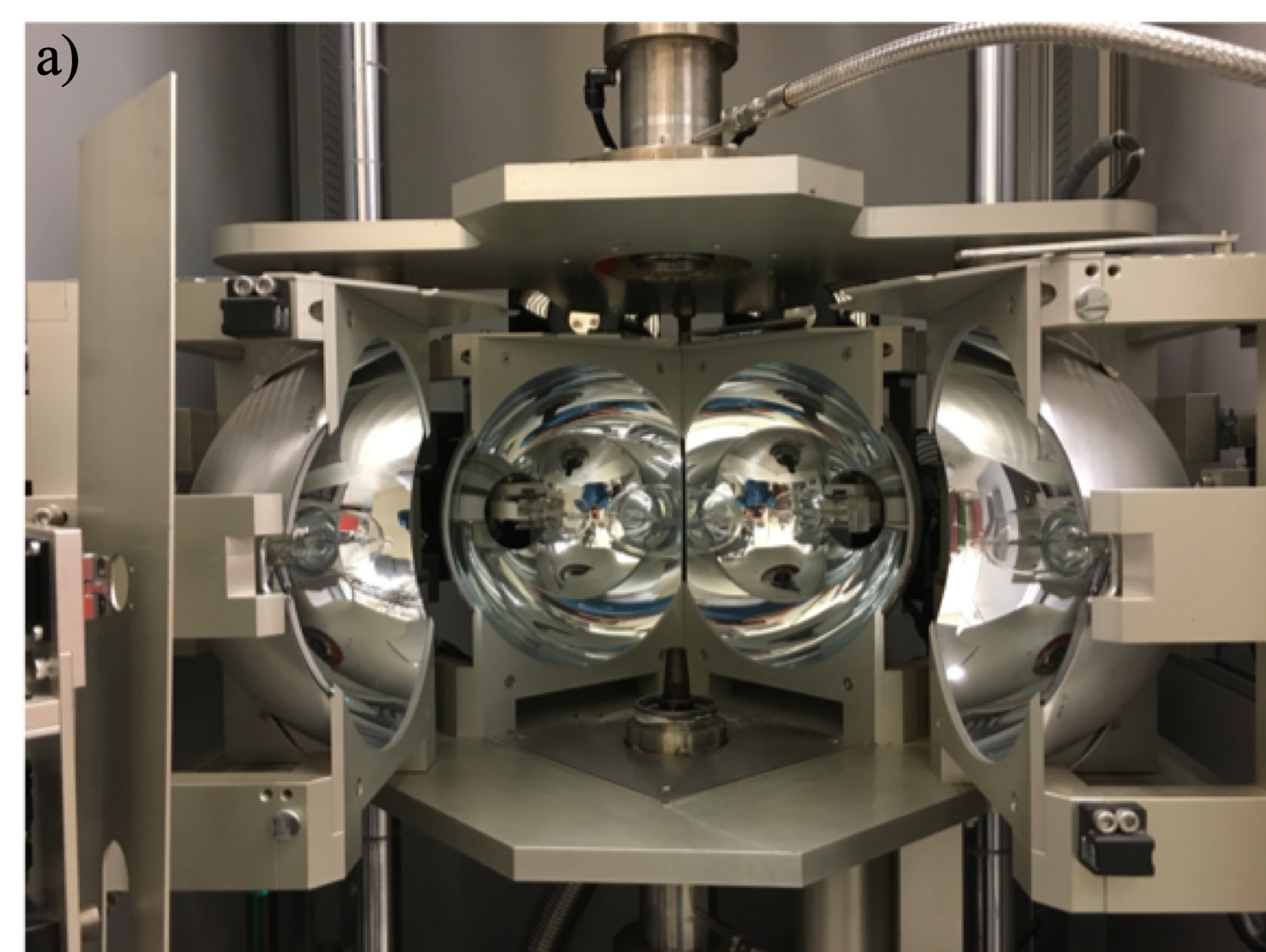}
    \includegraphics[width=0.3\textwidth]{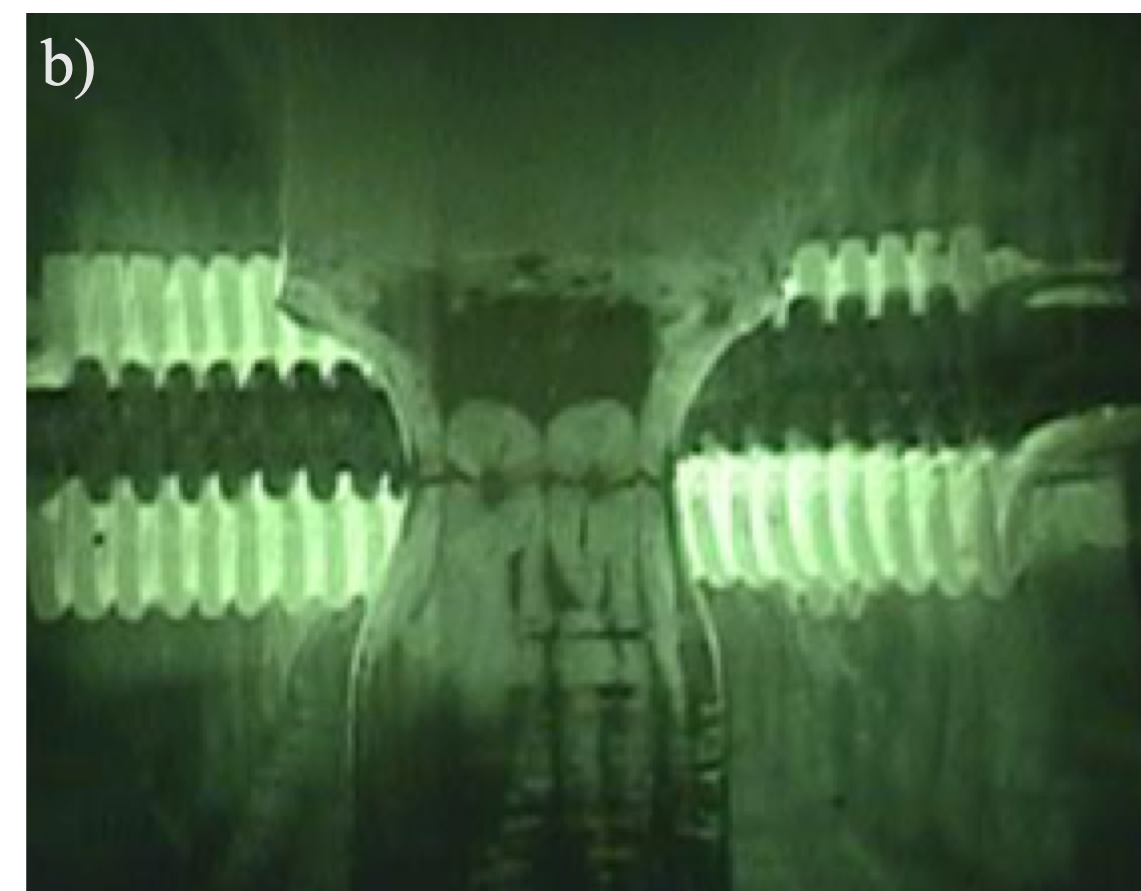}
    \caption{a. Infrared heated optical floating zone furnace and b. Image of molten zone between grown faceted crystal boule (bottom rod) and feedrod (top rod).}%
    \label{fig:IRFZ}%
\end{figure}

The floating zone technique has several advantages \cite{Feigelson2015} some of which are: it’s crucible free, it’s highly suited for incongruently melting materials (as high as $2500^oC$) due to the steep thermal gradient on the crystallization front, and it’s also an effective way to provide samples for the study of solid solutions and doped materials phase diagrams \cite{Baggari2018, Savitzky}. On the other hand, this technique has disadvantages if the materials growth involves very high vapor pressure – volatile species, low surface tension or high viscosity materials or if the stress leading to cracks in the cooling crystal are to be avoided. However, even in these case particular solutions can be implemented such as using a high pressure gas flow to suppress vaporization or promote oxidation, a traveling solvent to facilitate the melting/crystallization, or  annealing for reduction of stress immediately after the melt zone or once the growth is completed.  \\

The growth process starts with first preparing an x-ray pure polycrystalline sample by repeated mixing, grinding in mortar and pestle, pelletizing and annealing in high temperature furnace. Then strong  ceramic rods (feedrods) which are typically  8 mm diameter and 100 mm long are cold/hot pressed under about 70 MPa pressure to achieve highest density (usually between 50\% to 90\%). Sometimes, an initial pre-melting of the feedrod helps in the densification of the feedrods if evaporation is not an issue. The chemical purity, uniformity and density of the feedrod are highly correlated to the growth stability and quality of the resulting crystal.  \\

\begin{figure}[h]%
    \centering
    \includegraphics[width=0.35\textwidth]{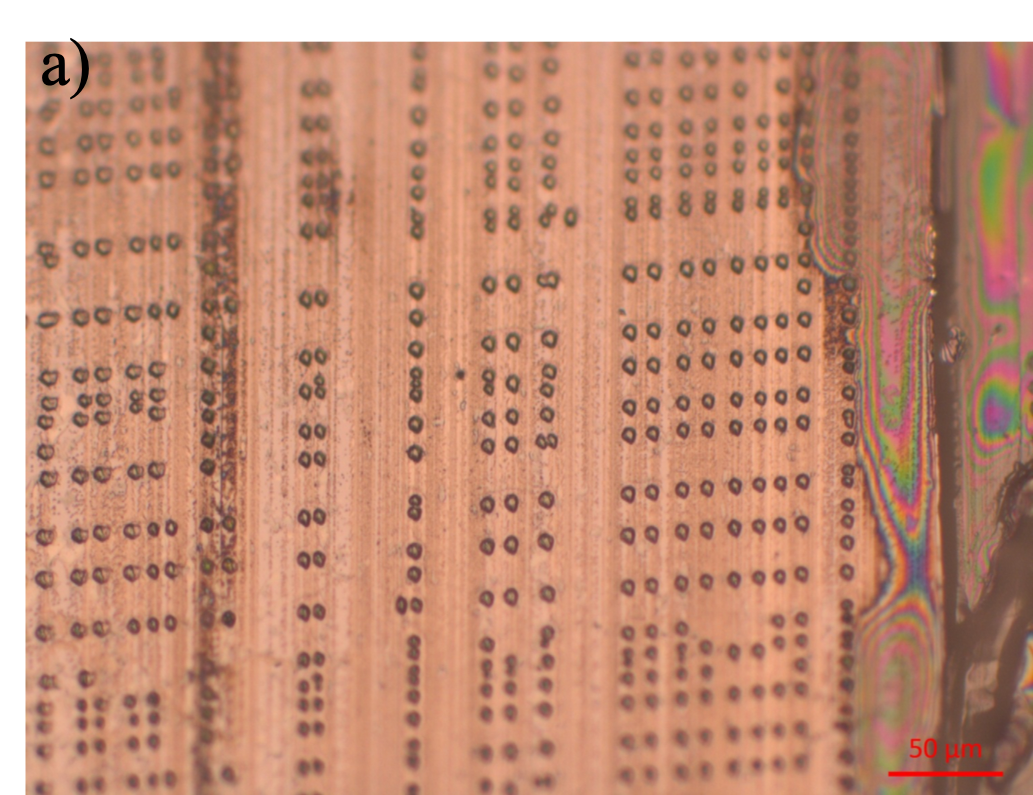}
    \includegraphics[width=0.4\textwidth]{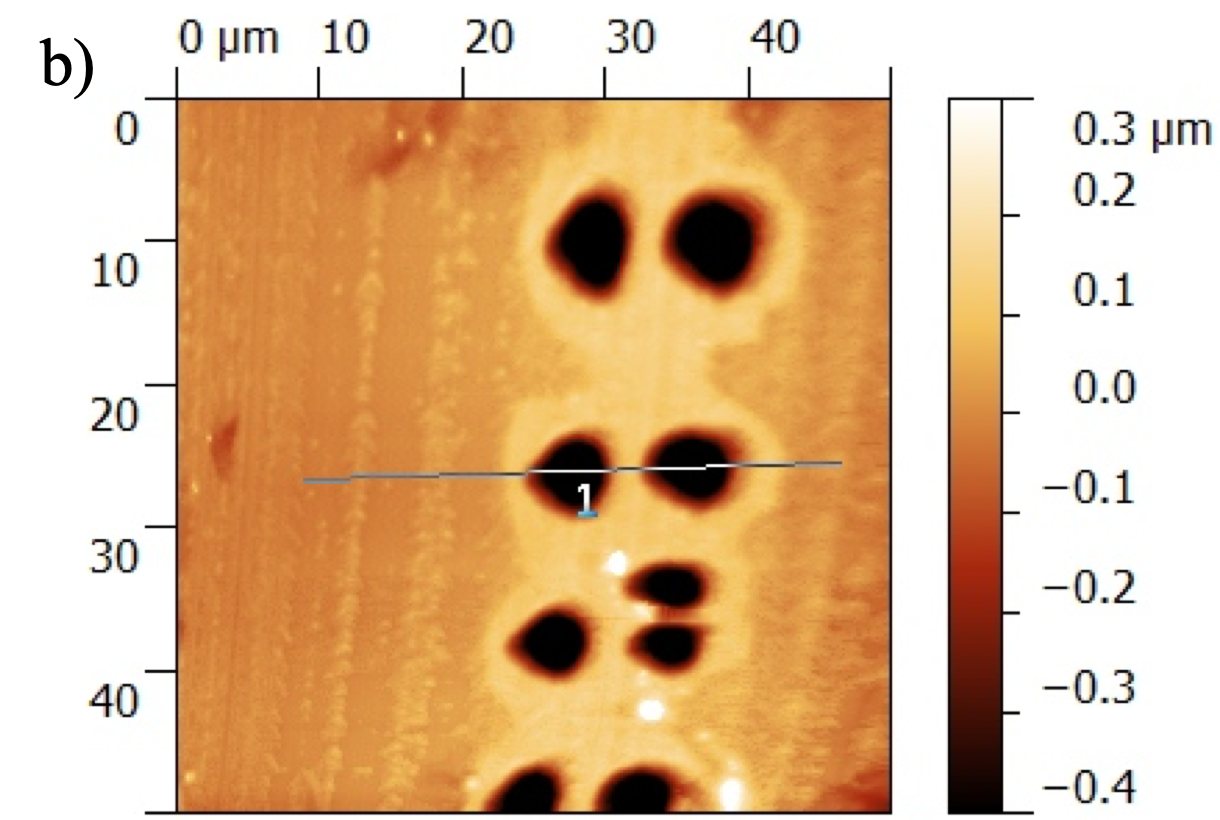}
    \includegraphics[width=0.4\textwidth]{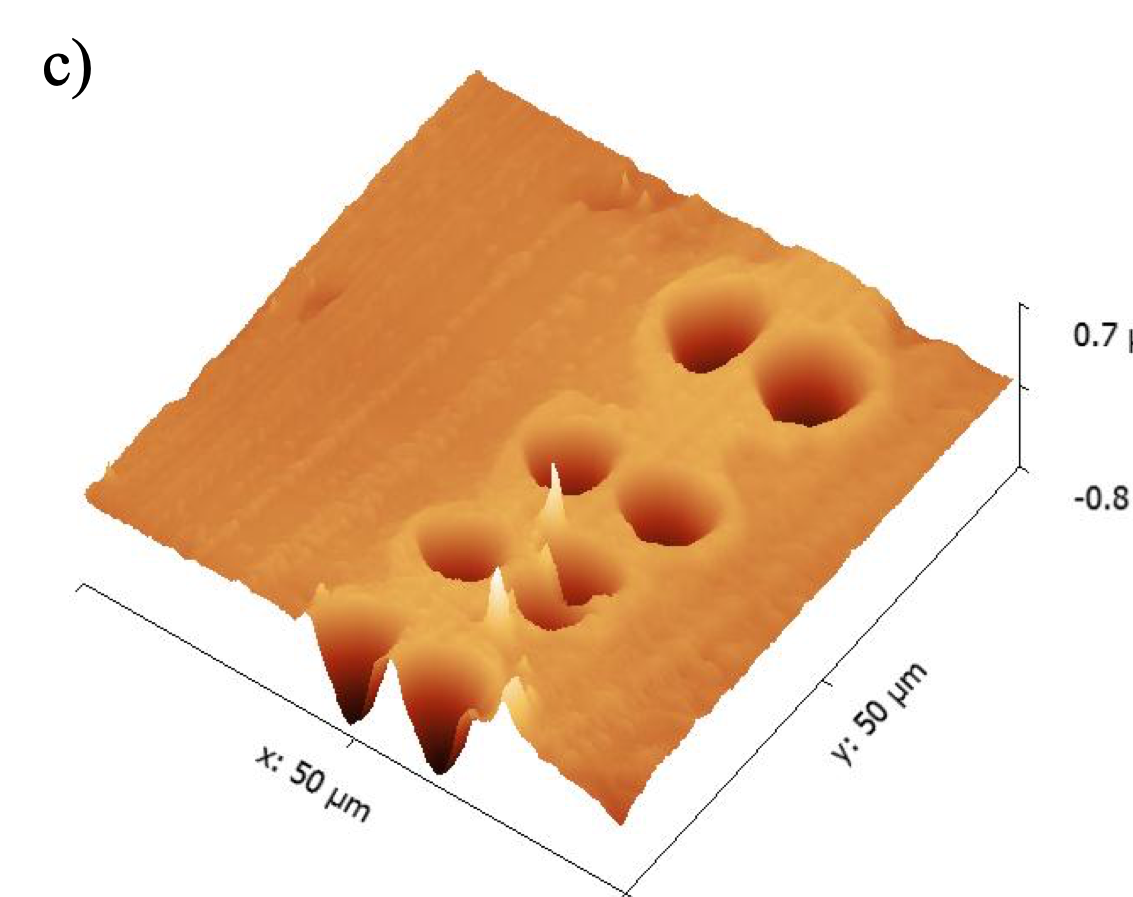}
    \caption{a. Microbubbles periodically present in the cleaved surface of IR-FZ grown crystal $PrCa_{1.8}Sr_{0.2}Mn_2O_7$ seen under a polarized optical microscope, b. and c. Atomic force microscope (AFM) images showing the surface topology, indicating the dots are holes on the cleaved surface, suspected from X-ray powder diffraction as arising from $MnO_2$ evaporation \cite{labfigures}.}%
    \label{fig:bubbles}%
\end{figure}

Once the feed and seed rods are appropriately mounted in a sealed quartz tube inside the FZ furnace, the feeding and opposite rotation speeds (used for uniformity / zone leveling) and growth power (temperature) are monitored through a video (CCD) camera in addition permitting a visual observation in situ of the crystal growth processes: such as the initial fusion and necking of the feedstock and seed, bubble formation inside the molten zone, zone collapse,  developing cracks (microbubbles) in the grown crystal etc. as in Figure \ref{fig:bubbles}.  \\ %amusingly known as Solomon's bubbles in the RCEM lab for first observing them.

While defects in FZ crystal growth can happen due to artificial sources such as rotation etc., compositional variations in the grown crystal can also be induced by the underlying process of thermal convection (so called Marangoni convection). Such temperature and flow inhomogeneities result in macro- and microscopic variations of the crystal compositions and commonly can visually be seen as “striations” of several hundred micrometers sizes on cleaved surfaces of the crystal. Several methods have been suggested to overcome the problem of striations – such as by controlling the flow velocity below a certain experimentally determined  critical value.  \\

It’s worthy to mention that gravity governs the shape of the liquid zone (bottle neck shape) in addition to surface tension. In ordinary FZ configurations, the maximum diameter of the melt zone is less than the zone length so that the feedrod and seed cannot be connected by a solid region. Hence, the maximum diameter of FZ crystals grown is 20 mm and if using additional stabilizing mechanisms such as in industrial applications, the diameter is rarely larger than 150 mm.   

%%%%%%%%%%%%%%%%%%%%%%%%%%%%%%%%%%%%%%

\subsubsection{Laser-diode Heated Floating Zone Method}

To improve upon the broad vertical light focusing and horizontal inhomogeneous heating that lead to growth instability and thermal degradation in the conventional Infrared lamp-heated floating zone (IR-FZ) method, a new variant of the floating-zone technique called the laser-diode-heated floating-zone (L-FZ) has been developed \cite{Ito2013} . In L-FZ technique, the molten-zone is established using a set of (five) horizontal laser beams unlike a conventional mirror-type FZ. Around the rotation axis of the FZ, equally spaced equivalent high power laser diodes are placed and cooperatively irradiate the molten zone by concentrated and homogeneous laser beams. The non-divergent property of the laser beam in L-FZ allows for the growth of high melting temperature refractory materials comprising volatile components such as $V_2O_3, In_2O_3$ \cite{Prachi, Li, Ozcelik} or alloys.  \\ 

In addition, the quality of the crystals is much improved with L-FZ as seen in the facets which appear often and from the clear solid-liquid interface which promotes stable crystal growth. The L-FZ technique has been shown to successively grow representatives of incongruently melting materials such as strongly correlated electron system $BiFeO_3$ and high-Tc superconductor $(La,Ba)_2CuO_4$, which were believed to be difficult to grow as single crystal by the conventional Infrared-heated FZ method.  \\

The L-FZ setup has several advantages compared to a conventional Infrared lamp mirror-type FZ furnace. The setup consisting of critical focusing of the laser beam permits a very sharp focus which results in temperature gradients exceeding 150 $^oC$/mm at the liquid-solid interface right outside the molten-zone and consequently high growth rates, as compared with other FZ procedures \cite{Prachi, Francisco, Ferreira}. This temperature gradient is much steeper compared to a conventional lamp heated mirror-type FZ furnace, where the gradient ranges from 20 to 40 $^oC$/mm depending on the heating geometry and power rating of lamps employed.   \\

\begin{figure}[h]%
    \centering
    \includegraphics[width=0.35\textwidth]{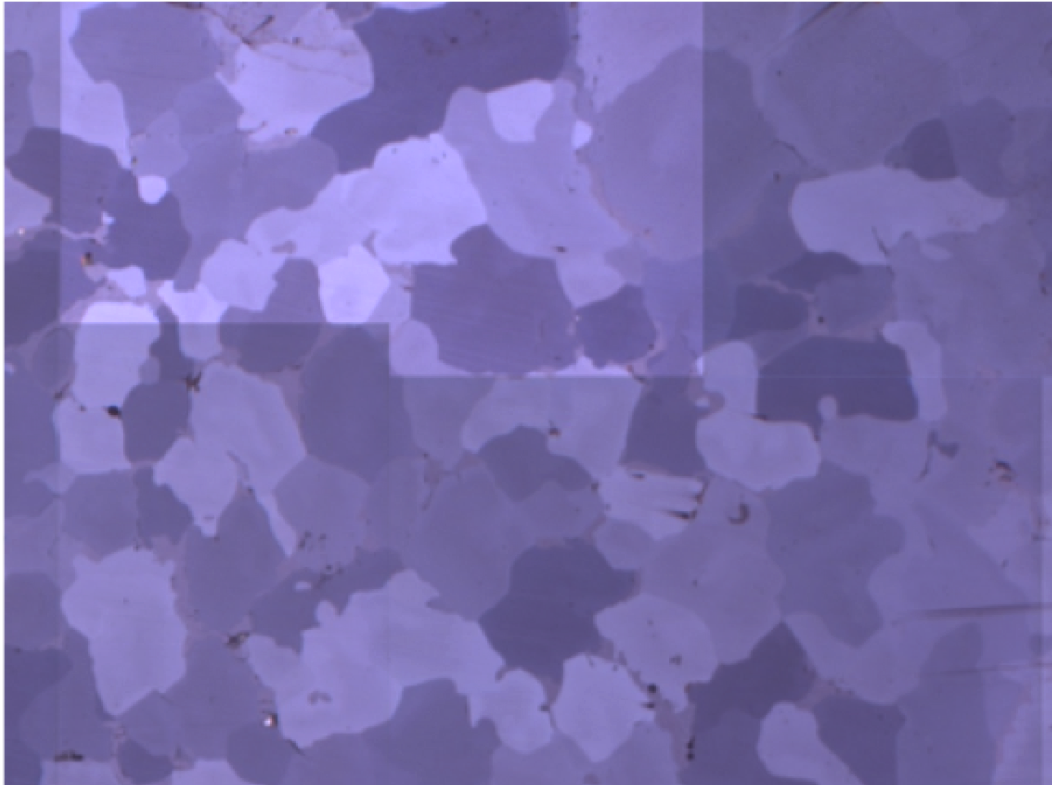}
    \caption{Cross-section of $GaFeO_3$ IR-FZ crystal under Polarized–light Optical Microscope. The low temperature gradient in conventional FZ furnaces can create constitutional supercooling where the growth becomes granular (grains diameter of approx. 100 $\mu$m).}%
    \label{fig:GFO}%
\end{figure}

When the temperature gradient along the rotation axis is low, the melt can attack the feed rod seriously and spill over to the crystal eventually making the growth unstable as the light is focused widely and heats the feed rod and the grown boule simultaneously. This is true even if traveling solvent floating zone (TSFZ) technique is used to lower the melting temperature of the FZ. Besides, the low temperature gradient in conventional FZ furnaces often creates constitutional supercooling where the growth becomes granular as shown in Figure \ref{fig:GFO} for IR-FZ grown $GaFeO_3$ .   \\

On the other hand, the presence of a very steep temperature gradient results in a shorter molten zone, decreasing the deleterious effect of convection currents in the liquid, with almost no evaporation from the region of the feed above the molten-zone and the shape of the grown material can be conformed by changing the drawing speed. Tolerance to the decentering of samples and highly efficient heating are also established in the L-FZ method. The homogeneous distribution of the heating power as well as that of the temperature along the rotational direction is observed to reduce defects in the grown crystals. Since the horizontal width of the homogeneous laser beam is larger than the diameter of the sample, stable growth is realized even if the feedrod and seed crystal maybe misaligned.  \\

In the present study, we use a laser-assisted image furnace from Crystal Systems Corporation, Japan installed at Rutgers University (USA) within the Rutgers Center for Emergent Materials. This L-FZ furnace is equipped with five semiconductor laser diodes with a total maximum power of 1 kW (5 $\times$ 200 W) as the heating source. The wavelength of the laser light is 976 nm $\pm$ 5 nm with a rectangular beam profile 4 mm (height) by 8mm (width). The pulling rate employed in the L-FZ growth can vary from 0.01mm/h to 300 mm/h, and the feed and seed rods can be counter rotated independently in the range 3 – 100 rpm. The growth chamber can sustain gas pressures up to 10 bar and also allows low pressure vacuum of up to 0.5 uTorr growth conditions. \\ 

%table
\begin{table}[h]
\begin{center}
    \begin{tabular}{ | p{2cm} | p{3cm} | p{3cm} |}
    \hline
\textbf{Property} &	\textbf{L-FZ (Crystal Systems, Inc.)}	& \textbf{IR-FZ (Quantum Design, Inc)} \\ \hline
Heating	 &  200W x 5 laser diodes    	& 650W x 2 halogen lamps \\ \hline
Temperature Homogeniety &	$>$  95 \%       & Not specified  \\ \hline
Maximum Temperature &	Not limited, melts sapphire with m.p. 2030 $^o$C easily &	2100 $^o$C \\ \hline
Temperature Gradient &	$>$ 150$^o$C/mm	& 30$^o$C/mm \\ \hline
Growth Speed &	0.01 – 300 mm/h	 & 0.1 – 14 mm/h \\ 
    \hline
    \end{tabular}
    \caption{Comparison between Laser-diode heated Floating Zone (L-FZ) and Infrared lamp heated Floating Zone (IR-FZ) furnaces from Crystal Systems, Inc. and Quantum Design, Inc respectively.}
    \label{table:FZtable}
\end{center}
\end{table}%

\begin{figure}[h]%
    \centering
    \includegraphics[width=0.35\textwidth]{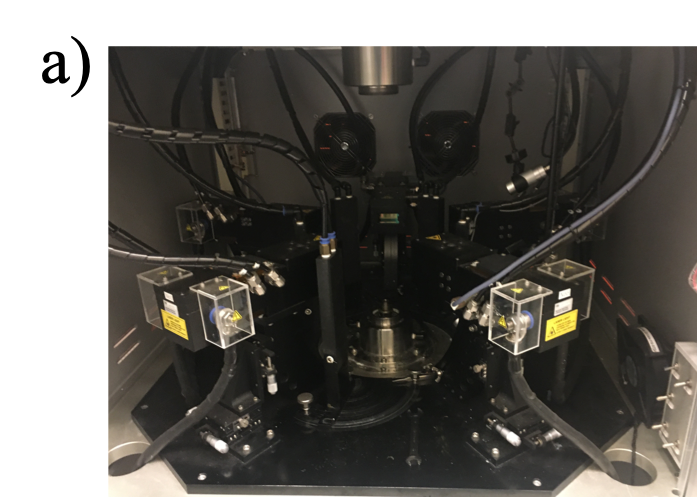}
    \includegraphics[width=0.35\textwidth]{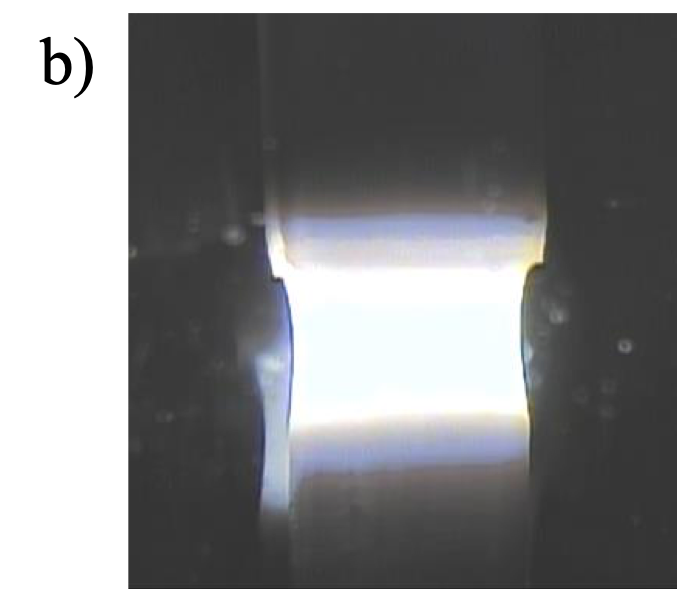}
    \caption{ a. Laser-diode heated floating zone furnace inside view and b. Image of molten zone between feedrod (top rod) and grown crystal boule (bottom rod) of $Tb_2SrAl_2O_7$.}%
    \label{fig:LFZ}%
\end{figure}

In our L-FZ experiments, the molten-zone length is usually about 3mm, which is less than or comparable to the diameter of the feed rod. The image of the region around the floating-zone during the L-FZ growth is continuously captured with a CCD camera. As can be seen in Figure \ref{fig:LFZ}, both melting (upper) and crystallization (lower) interfaces are very sharp and well-defined for typical growth. A strong evaporation from the molten-zone can occur for growths involving volatile components but an excellent feature of the L-FZ technique is that it can stabilize it and the high-temperature induced evaporation of the volatile component is not detrimental to the growth process as long as the amount lost is compensated by taking an appropriate excess of the volatile component in the feed rod. \\

It’s important to note that  while the use of 1.5kW halogen/Xenon lamps in a conventional mirror-type FZ can melt and form a workable floating-zone for most materials, an excessive evaporation from the molten-zone and the region of the feed immediately above the zone in the case of volatile components can lead to a gradual blackening of the upper portion of the quartz tube enclosure subsequently blocking the radiation and causing the growth experiment to often be prematurely terminated.   \\

The two techniques: conventional infrared lamp heated mirror-type FZ technique and the laser assisted FZ technique have essential difference in the heating mechanism. In the conventional mirror based FZ technique, due to a rather broad and inhomogeneous focusing, the light-flux arrives at the molten-zone after reflection from the mirror surface over a wide angular range depending on the eccentricity of the mirrors. On the other hand, in the L-FZ technique, the narrow laser beams in the plane reach the molten zone directly providing a remarkable stability to the molten-zone and a large-size boule as a result.

\section{Examples}

\subsubsection{The $RE_2SrAl(Sc)_2O_7$ Series}

Complex oxide systems containing mixed rare-earth and alkaline-earth metals attracted the particular attention of researchers owing to their possible use in electronic technology such as superconductors, ferroelectricity and colossal magnetoresistance \cite{Cheong2007, Ramesh2007}, catalytic activity etc. Among these complex oxides, layered perovskite-like compounds $RE_2SrAl_2O_7$, RE = Nd, Sm, Eu, Gd, Tb, Dy which belong to the Ruddlesden–Popper homologous series of general formula $SrRE_nMnO_{3n+1}$ with n=2, where M stands for Al or a 3d element have attracted special attention \cite{Zver2003, Titov2009}. The structure is a periodic stacking of a rock-salt layer and a double perovskite block in the sequence …( P )( P )( RS )( P )( P )( RS )… (the P2 / RS structure type). Single crystals and magnetic properties of the title refractory compounds are not known and only few electrical and thermal properties have been studied \cite{Feng2012}.  \\
% , see Figure  \ref{fig:p2rs}

We here report on the successful first bulk crystal growth of the phase-pure $RE_2SrAl_2O_7$, RE = Nd, Sm, Eu, Gd, Tb, Dy compounds in terms of single crystal syntheses paving the way for further study of their electrical and magnetic properties  as well as the understanding of n=2 Ruddlesden–Popper phases and their properties. The quality of the single crystals also offers the possibility to investigate the nature of frustrated magnetism in these series, analogously to the related pyrochlore compounds such as $Ho_2Ti_2O_7$ \cite{Fennell2009}. \\

% \subsection{Synthesis}

Ceramic sample of $RE_2SrAl_2O_7$, RE = Nd, Sm, Eu, Gd, Tb, Dy were prepared by direct solid-state reaction from stoichiometric mixtures of $RE_2O_3$, $SrCO_3$ and $Al_2O_3$ powders all from (Alfa Aesar, 99.99\%). The rare earth oxides were first dried at 800 $^oC$ overnight to remove any adsorbed $CO_2$ and $H_2O$. The stoichiometric mixtures were then thoroughly ground in an agate mortar and pestle and pelletized. The pellets were calcined at $900^oC$ in air for 12 hours and reground and heated at 1400 – 1530 $^oC$ for 20 h with intermediate grindings to ensure a total reaction similarly to \cite{Chun2010, Zver2003, Ska1999}.  \\

\begin{figure}[h]%
    \centering
    \includegraphics[width=0.4\textwidth]{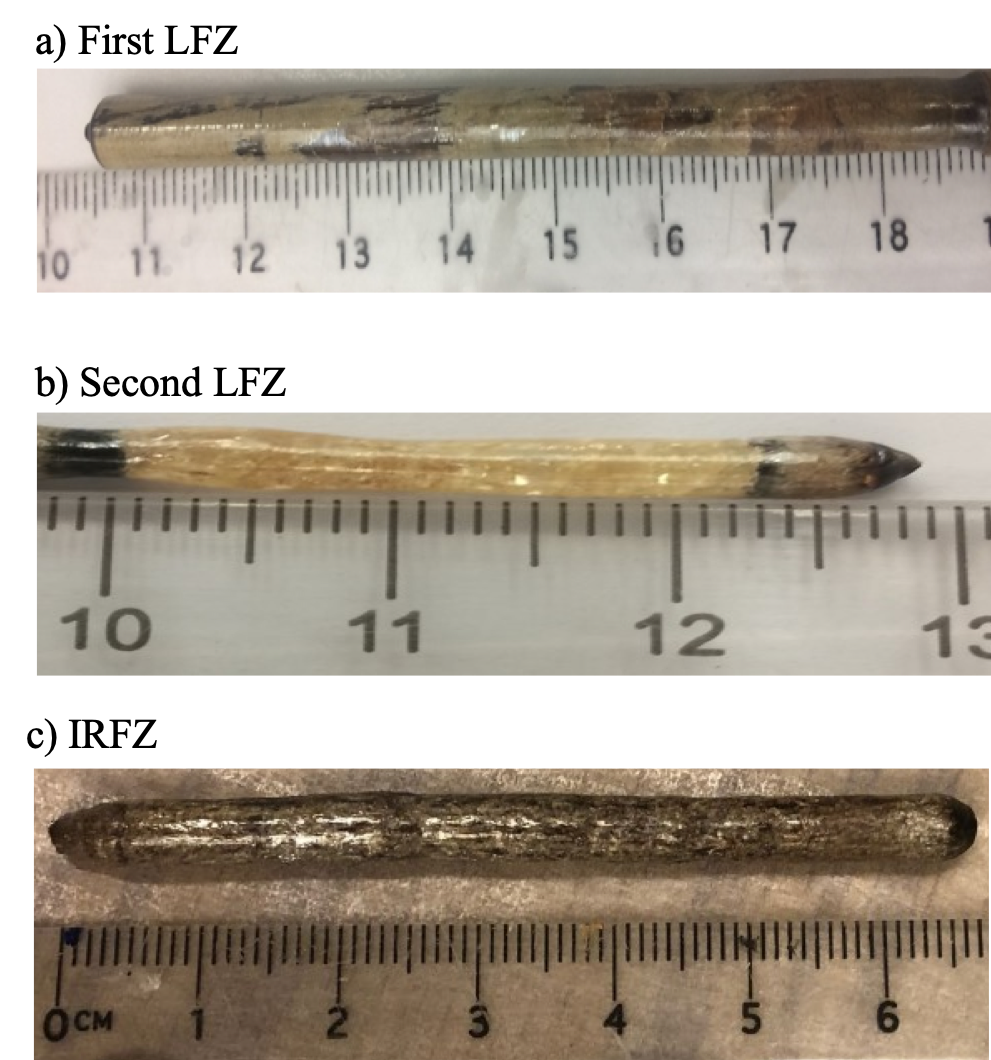}
    \caption{$Tb_2SrAl_2O_7$ crystal boules grown by L-FZ (a. and b.) and IR-FZ (c.) respectively. The transparent nature of two-step L-FZ grown boule indicates higher quality and more stable growth process as compared to the IR-FZ boule.}%
    \label{fig:boules}%
\end{figure}

Large single crystals of this refractory series were grown for the first time using a laser-diode heated floating zone technique explained above. Two step growth utilizing a previously grown crystal as seed was employed to obtain high crystal quality, shown in Figure \ref{fig:boules}) b as compared to a similarly IR-FZ grown non-transparent boule. The optimal growth conditions were growth speed of 2 - 4mm/hr, atmospheric air flow of 0.1 L/min and counter-rotation of the feed and seed rods at 20 and 30 rpm, respectively.   \\

% Figure 2.3: crystal pictures

It's important to also mention that a similar family of compounds $Ln_2SrSc_2O_7$ (Ln =  La - Tb) have been synthesized by solid-state reaction method and reported to crystallize in the Cmc21 space group which is a polar space group \cite{Sharits2016}. However, ferroelectric measurements on $La_2SrSc_2O_7$ and $Nd_2SrSc_2O_7$ bulk ceramic pellets showed no hysteresis in the polarization versus applied field or the polarization versus frequency suggesting these compounds are not ferroelectric.  \\

Within these rare-earth series, the competition and balance of RE-Sr ordering and the structure-bonding anisotropy also fixes the stability of the scandate to be considered. This is reflected in the considerable impurity encountered in the synthesis of $Ho_2SrSc_2O_7$ which suggests a limit of stability in the series. An analysis of the evolution of the stability of the P2/RS intergrowth and cation ordering in the scandates reveals a partially ordered arrangement Ln-Sr atoms in $Ln_2SrSc_2O_7$ (Ln = La-Sm) and a fully ordered arrangement for  Ln = Eu - Tb.  \cite{Titov2009, Zver2003,Seono92}.  \\

We synthesized ceramic sample of $RE_2SrSc_2O_7$, RE = Gd, Tb  by direct solid state reaction from stoichiometric mixtures of $RE_2O_3$, $SrCO_3$ and $Sc_2O_3$ powders all from (Alfa Aesar, 99.99\%) at 1350 – 1400 $^oC$ for 20 h with intermediate grindings as in \cite{Chun2010, Zver2003, Ska1999, Seono92, Titov2009}. Subsequently, large single crystals of this refractory series were grown for the first time using a laser-diode heated floating zone technique. The electrical and magnetic properties of these series will be reported in a subsequent publications. \\

\subsubsection{The Stuffed Tridymite $BaCoSiO_4$}

The tridymite structure is a high-temperature stable polymorph of silica ($SiO_2$) \cite{Buerger1954}. Tridymite derivatives such as $BaMnSiO_4$ \cite{Nihtianova1997}, $BaFeGaO_4$ \cite{Kahlen2002}, $BaAl_2O_4$ \cite{Abak2000}, $BaMSiO4$ (M = Co, Mg, Zn) \cite{Liu1993} have been studied for their rich physical properties (e.g. ferroelectricity \cite{Tanba1985, Zou2017}, photo-dielectric effect \cite{Tanigu2014}, stripe domain formation \cite{Yamamoto1992, Abak2000}, complex polymorphism and topology of tetrahedral framework \cite{Buerger1954, Tanaka2014, Kahlen2000,Hobbs2000}).  \\

A stuffed derivative, such as $BaCoSiO_4$, can occur when lower valence tetrahedrally coordinated ions (here Co$^{2+}$) partially replace Si$^{4+}$ ions and by stuffing Ba$^{4+}$ ions into the voids of the tridymite framework keeping the charge balance \cite{Liu1993}. Specifically, the crystal structure of $BaCoSiO_4$, similar to that of Kalsilite ($KAlSiO_4$), consists of a Ba-stuffed tetrahedral framework derived from that of $SiO_2$ tridymite with almost triangular shaped six-membered rings of corner-shared tetrahedra pointing alternately up and down and stacked along the c direction, joined via the O(4) oxygen atoms in a staggered configuration.  \\ %see Figure \ref{fig:bcsostructure2}

Tridymite derivatives such as $BaMnSiO_4$ \cite{Nihtianova1997}, $BaFeGaO_4$ \cite{Kahlen2000}, $BaAl_2O_4$ \cite{Abak2000}, $BaMSiO_4$ (M = Co, Mg, Zn) \cite{Liu1993} have been studied for their rich physical properties (e.g. ferroelectricity \cite{Naomi1985,Zou2017}, photo-dielectric effect \cite{Tanigu2014}, stripe domain formation \cite{Abak2000,Yamamoto1992}, complex polymorphism and topology of tetrahedral framework \cite{Buerger1954, Tanaka2014}). In the family of known Ba-stuffed tridymite derivative compounds, the silicates $BaMSiO_4$ ($M = Mg, Zn, Co$) and the germanate $BaZnGeO_4$  are the only ones known to crystallize with a ($\sqrt{3}xA, C$) superstructure of the kalsilite type, while $BaAl2O_4$ adopts a kalsilite-like topology but with a different superstructure of ($2xA, C$) supercell. \\

$BaCoSiO_4$ crystalizes at room temperature in the polar space group $P6_3$ and its structure, similar to that of Kalsilite ($KAlSiO_4$), consists of a Ba-stuffed tetrahedral framework derived from that of $SiO_2$ tridymite with almost triangular shaped six-membered rings of corner-shared tetrahedra pointing alternately up and down with stacking along the c-direction in a staggered configuration. The Co and Si atoms are completely ordered, with $BaCoSiO_4$ having the highest degree of tetrahedral tilting among $BaMSiO_4$ ($M = Mg, Zn, Co$) \cite{Liu1993} prompting to synthesize single crystal of $BaCoSiO_4$ for further investigation of these compounds.  \\ %see Figure \ref{fig:bcsostructure2}

\begin{figure}[h]%
    \centering
    \includegraphics[width=0.35\textwidth]{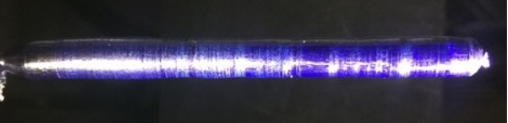}
    \caption{A 60 mm boule of L-FZ grown $BaCoSiO_4$. Unlike most FZ grown crystal boules, $BaCoSiO_4$ cleaves easily in the ab plane with cracks appearing perpendicular to the growth direction shown here. }%
    \label{fig:bcsoxrdboule2}%
\end{figure}

We here report synthesis of bulk $BaCoSiO_4$ single crystals by the floating zone method. Powder sample of $BaCoSiO_4$ was prepared by direct solid state reaction from stoichiometric mixtures of $BaCO_3$, $Co_3O_4$ and $SiO_2$ powders all from (Alfa Aesar, 99.99\%) as previously reported \cite{Liu1993, Tanigu2014}. The mixture was calcined at 1150 K in air for 12 hours and then re-ground, pelletized and heated at 1450 K for 20 h and at 1500 K for 15 h with intermediate grindings to ensure a total reaction. The resulting powder sample is very fine and bright blue in color. Small hexagonal flux crystals ($3mm^2 \times 11mm$) can be grown from NaCl/KCl flux or by slow cooling (2 K/hr) from the melt in Alumina Crucible.  Large single crystals, see Figure \ref{fig:bcsoxrdboule2} b), were grown using a laser-diode heated floating zone technique. The optimal growth conditions were growth speed of 2-4mm/hr, atmospheric air flow of 0.1 L/min and counter-rotation of the feed and seed rods at 15 and 30 rpm, respectively.  \\

\section{Conclusion}

We reviewed current state of the art in the floating zone method with particular emphasis on the Laser-diode heated floating zone method. We demonstrated the methods with synthesis results of high-quality single crystals of the refractory $n=2$ Ruddlesden–Popper homologous series $RE_2SrAl(Sc)_2O_7$ and the stuffed tridymite-type $BaCoSiO_4$. The ability to further attain full capability to control complex atomic- and subatomic-level interactions so that new or artificial forms of matter that have specific, tailored properties can be realized requires a new approach in crystal fabrication methods including developing sophisticated automated single crystal synthesis optimization and characterization tools and instruments. As such new highly-controlled computerized techniques should be adopted for material discovery, design and manufacture in the field to help create materials-based technologies significantly faster. \\
% \vspace{2mm} %5mm vertical space

\textbf{Acknowledegments}
The experimental work was performed at the Rutgers Center for Emergent Materials and the Department of Physics and Astronomy.  D.S.V. would like to acknowledge United States-India Educational Foundation (USIEF) for Fulbright-Nehru Doctoral Fellowship.  The authors thank Prof. Sang-Wook Cheong for the advice on the methods demonstrated here. \\

% Part of this work is published in the first author's thesis.

%put thebibliography in s a seprate tex and import here?
% \import{}{bib}   %uncomment as an alternative to below if needed

% \begin{thebibliography}

% % \bibitem{latexcompanion} 
% % Michel Goossens, Frank Mittelbach, and Alexander Samarin. 
% % \textit{The \LaTeX\ Companion}. 
% % Addison-Wesley, Reading, Massachusetts, 1993.
% \end{thebibliography} 


%\nocite{*}
%\bibliography{mainscript}% Produces the bibliography via BibTeX.

\begin{thebibliography}{44}%
\makeatletter
\providecommand \@ifxundefined [1]{%
 \@ifx{#1\undefined}
}%
\providecommand \@ifnum [1]{%
 \ifnum #1\expandafter \@firstoftwo
 \else \expandafter \@secondoftwo
 \fi
}%
\providecommand \@ifx [1]{%
 \ifx #1\expandafter \@firstoftwo
 \else \expandafter \@secondoftwo
 \fi
}%
\providecommand \natexlab [1]{#1}%
\providecommand \enquote  [1]{``#1''}%
\providecommand \bibnamefont  [1]{#1}%
\providecommand \bibfnamefont [1]{#1}%
\providecommand \citenamefont [1]{#1}%
\providecommand \href@noop [0]{\@secondoftwo}%
\providecommand \href [0]{\begingroup \@sanitize@url \@href}%
\providecommand \@href[1]{\@@startlink{#1}\@@href}%
\providecommand \@@href[1]{\endgroup#1\@@endlink}%
\providecommand \@sanitize@url [0]{\catcode `\\12\catcode `\$12\catcode
  `\&12\catcode `\#12\catcode `\^12\catcode `\_12\catcode `\%12\relax}%
\providecommand \@@startlink[1]{}%
\providecommand \@@endlink[0]{}%
\providecommand \url  [0]{\begingroup\@sanitize@url \@url }%
\providecommand \@url [1]{\endgroup\@href {#1}{\urlprefix }}%
\providecommand \urlprefix  [0]{URL }%
\providecommand \Eprint [0]{\href }%
\providecommand \doibase [0]{https://doi.org/}%
\providecommand \selectlanguage [0]{\@gobble}%
\providecommand \bibinfo  [0]{\@secondoftwo}%
\providecommand \bibfield  [0]{\@secondoftwo}%
\providecommand \translation [1]{[#1]}%
\providecommand \BibitemOpen [0]{}%
\providecommand \bibitemStop [0]{}%
\providecommand \bibitemNoStop [0]{.\EOS\space}%
\providecommand \EOS [0]{\spacefactor3000\relax}%
\providecommand \BibitemShut  [1]{\csname bibitem#1\endcsname}%
\let\auto@bib@innerbib\@empty
%</preamble>





\bibitem{elsevier_crystal_growth} Progress in Crystal Growth and Characterization of Materials https://www.journals.elsevier.com/

\bibitem{Dong2015} S Dong, JM Liu, SW Cheong, Z Ren, Multiferroic materials and magnetoelectric physics: symmetry, entanglement, excitation, and topology, Advances in Physics  64 (5-6), 519-626 (2015)

\bibitem{Feigelson2015} Robert S. Feigelson, Crystal Growth through the Ages: A Historical Perspective, Geballe Laboratory For Advanced Materials, Stanford University, Stanford, CA, USA, (2015)

\bibitem{Tatarchenko2003} Y. A. Tatarchenko, Shaped Crystal Growth, Centre for Nuclear Energy, Grenoble, France (2003)

\bibitem{Theuerer1952} Theurer, Henry C. Method of Processing Semiconductive Materials, U. S. Patent 3,060,123 (Filed December 17, 1952. Issued October 23, 1962)

\bibitem{Jin} Jin, W., Drueke, E., Li, S., Admasu, A., Owen, R., Day, M., Sun, K., Cheong, S.W. and Zhao, L., 2020. Observation of a ferro-rotational order coupled with second-order nonlinear optical fields. Nature Physics, 16(1), pp.42-46.

\bibitem{Chen} Chen, C., Kim, H. S., Admasu, A. S., Cheong, S. W., Haule, K., Vanderbilt, D.,  Wu, W. (2018). Trimer bonding states on the surface of the transition-metal dichalcogenide TaT e 2. Physical Review B, 98(19), 195423.



\bibitem{Baggari2018} El Baggari, I., Savitzky, B. H., Admasu, A. S., Kim, J., Cheong, S. W., Hovden, R.,  Kourkoutis, L. F. (2018). Nature and evolution of incommensurate charge order in manganites visualized with cryogenic scanning transmission electron microscopy. Proceedings of the National Academy of Sciences, 115(7), 1445-1450.

\bibitem{Savitzky} Savitzky, B. H., El Baggari, I., Clement, C. B., Waite, E., Goodge, B. H., Baek, D. J., ..., Kourkoutis, L. F. (2018). Image registration of low signal-to-noise cryo-STEM data. Ultramicroscopy, 191, 56-65.

\bibitem{labfigures} Figures from unpublished manuscript. Courtesy of Sang-Wook Cheong.

\bibitem{Fennell2009} Fennell, Tom, P. P. Deen, A. R. Wildes, K. Schmalzl, D. Prabhakaran, A. T. Boothroyd, R. J. Aldus, D. F. McMorrow, and S. T. Bramwell. Magnetic Coulomb phase in the spin ice $Ho_2Ti_2O_7$. Science 326, no. 5951: 415-417(2009)

\bibitem{Chun2010} C. Wan, T. D. Sparks, P. Wei and D. R. Clarke, Thermal Conductivity of the Rare‐Earth Strontium Aluminates, J. Am. Ceram. Soc., 93 [5] 1457–1460 (2010)


\bibitem{Ito2013} T. Ito, T. Ushiyama, Y. Yanagisawa, Y. Tomioka, I. Shindo and A. Yanase, Growth of Highly Insulating Bulk Single Crystals of Multiferroic BiFeO3 and Their Inherent Internal Strains in the Domain-Switching Process, Journal of Crystal Growth Volume 363, 264-269 (2013)

\bibitem{Li} Li, Yanbin and Wang, Yazhong and Tan, Wenjie and Wang, Wenbo and Zhang, Junjie and Kim, Jae Wook and Cheong, Sang-Wook and Tao, Xutang; Laser floating zone growth of improper geometric ferroelectric $GdInO_3$ single crystals with Z6 topological defects; J. Mater. Chem. C, 6, 26 7024-7029, The Royal Society of Chemistry (2018)

\bibitem{Prachi} Prachi Telang, Andrey Maljuk, Dibyata Rout, Rongwei Hu, Markos Skoulatos, Koushik Karmakar, Silvia Seiro, Bertrand Roessli, Uwe Stuhr, Bernd Büchner, Sang-Wook Cheong, Surjeet Singh, Laser-diode-heated floating-zone crystal growth of $ErVO_3$,Journal of Crystal Growth, Volume 507, Pages 406-412 (2019)

\bibitem{Ozcelik} B. Ozcelik, G. Çetin, M. Gürsul, M.A. Madre, A. Sotelo, S. Adachi, Y. Takano, Low temperature thermoelectric properties of K-substituted $Bi_2Sr_2Co_2O_y$ ceramics prepared via laser floating zone technique, Journal of the European Ceramic Society, Volume 39, Issue 10, 3082-3087 (2019)

\bibitem{Francisco} Francisco Rey-García, Carmen Bao-Varela and Florinda M. Costa. Laser Floating Zone: General Overview Focusing on the Oxyorthosilicates Growth, Intech Open, (2019).

\bibitem{Ferreira} N.M. Ferreira, N.R. Neves, M.C. Ferro, M.A. Torres, M.A. Madre, F.M. Costa, A. Sotelo, A.V. Kovalevsky, Growth rate effects on the thermoelectric performance of $CaMnO_3$-based ceramics, Journal of the European Ceramic Society, Volume 39, Issue 14, 4184-4188 (2019)

\bibitem{Cheong2007} S.-W. Cheong and M. Mostovoy; Multiferroics: a magnetic twist for ferroelectricity, Nature Mater. 6:13-20 (2007)

\bibitem{Ramesh2007} R. Ramesh and N.A. Spaldin.; Multiferroics: progress and prospects in thin films , Nat Mater. 6:21-9 (2007)

\bibitem{Zver2003} I.Zvereva, Yu. Smirnov, V. Gusarov, V. Popova and J. Choisnet, Complex aluminates $RE_2SrAl_2O_7$ (RE = La, Nd, Sm-Ho): Cation ordering and stability of the double perovskite slab-rocksalt layer P2/RS intergrowth,  Solid State Sci. 5  343-9 (2003)

\bibitem{Ska1999} J. M. S. Skakle and R. Herd; Crystal chemistry of $(RE,A)_2M_3O_7$ compounds (RE=Y, lanthanide; A=Ba, Sr, Ca; M=Al, Ga), Powder Diffraction 14 (3) (1999)

% \bibitem{Zver2001} Zvereva I, Smirnov Y, Choisnet J., Prominent part of calcium ordering in the formation and stability of the intergrowth type solid solution $La_2Sr_{1−x}Ca_xAl_2O_7$, Int J Inorg Mater.3(1):95-100 (2001). 

\bibitem{Seono92} In-Seon Kim, Hitoshi Kawaji, Mitsuru Itoh, Tetsurō Nakamura, Structural and dielectric studies on the new series of layered compounds, strontium lanthanum scandium oxides, Materials Research Bulletin, Volume 27, Issue 10, 1193-1203 (1992)

\bibitem{Titov2009} Y. A. Titov, N. M. Belyavina, V. Ya. Markiv,M. S. Slobodyanik, Ya. A. Krayevska, V. V. Chumak; Reports of the National Academy of Sciences of Ukraine, 3:155-161 (2009)

\bibitem{Kahlen2000} V. Kahlenberg, R. X. Fischer and J. B. Parise, The Crystal Structures of the Strontium Gallates $Sr_{10}Ga_6O_{19}$ and $Sr_3Ga_2O_6$,  Journal of Solid State Chemistry 154, 612-618 (2000)


\bibitem{Abak2000}  A. M. Abakumov, O. I. Lebedev, L. Nistor , G. Van Tendeloo and S. Amelinckx, The ferroelectric phase transition in tridymite type $BaAL2O4$ studied by electron microscopy, Phase transitions 71,  143-160 (2000)

\bibitem{Liu1993}  B. Liu and J. Barbier, Structures of the Stuffed Tridymite Derivatives, $BaMSiO_4$ (M = Co, Zn, Mg), Journal of Solid State Chemistry 102, (1993) 115-125. 

\bibitem{Tanba1985}  N. Tanb, M. Wada and Y. Ishibashi, Soft Mode in the Ferroelectric Phase Transition of $Li_2Ge_7O_{15}$,  J. Phys. Soc. Jpn. 54,  4783-4786 (1985)


\bibitem{Zou2017}  Z-Y Zou, Z-H Chen, X-K Lan, W-Z Lu, B. Ullah, X-H Wang, Wen Lei, Weak ferroelectricity and low-permittivity microwave dielectric properties of $Ba2Zn_{(1+ x)}Si_2O_{(7+ x)}$ ceramics, Journal of the European Ceramic Society 37 3065–3071  (2017)

\bibitem{Tanigu2014}  H. Taniguchi, H. Moriwake, A. Kuwabara, T. Okamura, T. Yamamoto, R. Okazaki, M. Itoh, and I. Terasaki, Photo-induced change of dielectric response in $BaCoSiO_4$ stuffed tridymite,  J. Appl. Phys. 115, 164103 (2014).


\bibitem{Yamamoto1992}  N. Yamamoto, M. Kikuchi, T. Atake, A. Hamano, and Yasutoshi Saito, Incommensurate-Commensurate Phase Transition of $BaZnGeO_4$ Studied by Electron Microscopy and Diffraction, J. Phys. Soc. Jpn. 61, pp. 3178-3188 (1992)


\bibitem{Hobbs2000} Hobbs L.W., Yuan X. Topology And Topological Disorder In Silica. In: Pacchioni G., Skuja L., Griscom D.L. (eds) Defects in SiO2 and Related Dielectrics: Science and Technology. NATO Science Series, vol 2. Springer, Dordrecht  (2000)

\bibitem{Sharits2016} A. R. Sharits, Structure-Property Relationships in Noncentrosymmetric Layered Perovskites, Ohio State University, (2016)


\bibitem{Feng2012} J. Feng, B. Xiao, R. Zhou, W. Pan and D. R. Clarke, Anisotropic elastic and thermal properties of the double perovskite slab-rock salt layer $Ln_2SrAl_2O_7$ (Ln = La, Nd, Sm, Eu, Gd or Dy) natural superlattice structure, Acta Materialia 60  3380–3392 (2012)

\bibitem{Buerger1954}   M. J. Buerger, The stuffed derivatives of the silica structures , American Mineralogist 39 (7-8): 600-614 (1954).

\bibitem{Nihtianova1997}  D. D. Nihtianova, I. T. Ivanov, J. J. Macicek and I. K. Georgieva, Crystallographic data for $BaMnSiO4$: A new phase in the system $BaO-MnO-SiO_2$, Powder Diffraction 12(3)  167-170 (1997)

\bibitem{Kahlen2002} V. Kahlenberg, J. B. Parise, Y. Lee and A. TripathiIII, Characterization of the stuffed framework structures $BaAlGaO_4$ and $BaFeGaO_4$, Z. Kristallogr. 217  249-255 (2002)

\bibitem{Tanaka2014}  E. Tanaka, Y Ishii, H. Tsukasaki, H. Taniguchi and Shigeo Mori, Structural changes and microstructures in stuffed tridymite-type compounds $Ba_{1-x}Sr_{x}Al_{2}O_{4}$, Japanese Journal of Applied Physics 53, 09PB01 (2014)




\end{thebibliography}
%\bibliographystyle{apsrev}

%****************************texts below is for bibliography************************

\providecommand{\noopsort}[1]{}\providecommand{\singleletter}[1]{#1}%

\end{document}